\documentclass[10pt,a4paper]{article}

\usepackage{graphicx}
\usepackage{amssymb}
\usepackage{amsmath}          
\usepackage[latin1]{inputenc} 
\usepackage[T1]{fontenc}
\usepackage{float}            

\begin{document}

\title{\textbf{A New Causal Interpretation of EPR-B Experiment}}
      \author{
      \centering		
      \begin{tabular}{cc}
         Michel Gondran                                                  &Alexandre Gondran\\			
         \footnotesize\textit{University Paris Dauphine, Paris, France,} &\footnotesize\textit{SeT Lab, UTBM, Belfort, France,}\\
         \footnotesize\texttt{michel.gondran@polytechnique.org}          &\footnotesize\texttt{alexandre.gondran@utbm.fr}\\			
      \end{tabular}
}
\maketitle
 
\begin{abstract}
In this paper we study a two-step version of EPR-B experiment, the
Bohm version of the Einstein-Podolsky-Rosen experiment. Its
theoretical resolution in space and time enables us to refute the
classic "impossibility" to decompose a pair of entangled atoms
into two distinct states, one for each atom. We propose a new
causal interpretation of the EPR-B experiment where each atom has
a position and a spin while the singlet wave function verifies the
two-body Pauli equation. In conclusion we suggest a physical
explanation of non-local influences, compatible with Einstein's
point of view on relativity.
\newline
\newline
\textbf{keywords:} EPR-B - causal interpretation - entangled atoms - two-body Pauli equation - singlet state
\end{abstract}







\section{Introduction}

The nonseparability is one of the most puzzling aspects of quantum
mechanics. For over thirty years, the EPR-B, the spin version
proposed by Bohm~\cite{Bohm_1951,Bohm_1957} of the
Einstein-Podolsky-Rosen experiment~\cite{EPR}, the Bell
theorem~\cite{Bell64} and the BCHSH
inequalities~\cite{Bell64,CHSH,Bell_1987} have been at the heart
of the debate on hidden variables and non-locality; but hitherto
the precise nature of the physical process that lies behind the
"non-local" correlations in the spins of the particles has
remained unclear.

Many experiments since Bell's paper have demonstrated violations
of these inequalities and have vindicated quantum
theory~\cite{Clauser_1972,Fry_1976,Lamehi_1976,Aspect_1982a,Aspect_1982b,Tittel_1998,Zeilinger_1998,Beige,Rowe,Zeilinger_2002,Genovese_2005}.
The first one was done with pairs of entangled photons and clearly
violate Bell's
inequality~\cite{Aspect_1982a,Aspect_1982b,Tittel_1998,Zeilinger_1998}.
Entangled protons have also been studied in an early
experiment~\cite{Lamehi_1976}. The generation of EPR pairs of
massive atoms instead of massless photons has been
considered~\cite{Beige,Rowe}; it also shows experimental violation
of Bell's inequality with efficient detection~\cite{Rowe}.

In a new experiment, Zeilinger and all~\cite{Zeilinger_2007}
measure previously untested correlations between two entangled
photons, they show that these correlations violate an inequality
proposed by Leggett for non-local realistic
theories~\cite{Leggett_2003}.

The usual conclusion of these experiments is to reject the
non-local realism because the impossibility to decompose a pair of
entangled atoms into two states, one for each atom.

In this paper we show, on the EPR-B experiment, that this
decomposition is possible: a causal interpretation exists where
each atom has a position and a spin while the singlet wave
function verifies the two-body Pauli equation.

To demonstrate this; we consider a two-step version of EPR-B
experiment and we use an analytic expression of the wave function
and the probability density. The explicit solution is obtained via
a complete integration of the two-body Pauli equation \textit{over
time and space}.

A first causal interpretation of EPR-B experiment was proposed in
1987 by Dewdney, Holland and
Kyprianidis~\cite{Dewdney_1987b,Holland_1988b}. This
interpretation had a flaw: the spin of each particle depends directly on the singlet wave function, 
and so the spin module of each particle varied
during the experiment from 0 to $\frac{\hbar}{2}$.

The explicit solution in terms of two-body Pauli spinors and the
probability density for the two steps of the EPR-B experiment are
presented in section 2. The solution in space and time shows how
it is possible to deduce tests on the spatial quantization of
particles, similar to those of the Stern and Gerlach experiment.

In section 3, we provide a realistic explanation of the entangled
states and a method to desentangle the wave function of the two
particles.

The resolution in space of the equation Pauli is essential: it
enables the spatial quantization in section 2 and explains
determinism and desentangling in section 3.

In conclusion we propose a physical explanation of non-local
influences, compatible with Einstein's point of view on
relativity.

\section{Simulation and tests of EPR-B experiment in two steps}

\begin{figure*}
\begin{center}
\includegraphics{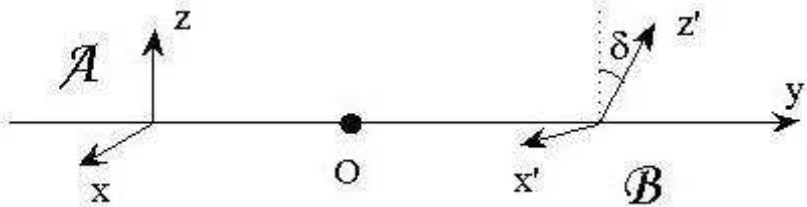}
\end{center}
\caption{\label{fig:expEPR}Schematic configuration of EPR-B
experiment.}
\end{figure*}

Fig.\ref{fig:expEPR} presents the Einstein-Podolsky-Rosen-Bohm
experiment. A source $S$ created in O pairs of identical atoms A
and B, but with opposite spins. The atoms A and B split following
0y axis in opposite directions, and head towards two identical
Stern-Gerlach apparatus $\mathcal{A}$ and $\mathcal{B}$.

The electromagnet $\mathcal{A}$ "measures" the A spin in the
direction of the Oz-axis and the electromagnet $\mathcal{B}$
"measures" the B spin in the direction of the Oz'-axis, which is
obtained after a rotation of an angle $\delta$ around the Oy-axis.

We further consider that atoms A and B may be represented by
Gaussian wave packets in x and z. We note $\textbf{r}= (x,z)$. The
initial wave function of the entangled state is the singlet state:
\begin{equation}\label{eq:7psi-1}
    \Psi_{0}(\textbf{r}_A,\textbf{r}_B) =\frac{1}{\sqrt{2}}f(\textbf{r}_A) f(\textbf{r}_B)(|+_{A}\rangle |-_{B}\rangle - |-_{A}\rangle | +_{B}\rangle)
\end{equation}
where $f(\textbf{r})=(2\pi\sigma_{0}^{2})^{-\frac{1}{2}}
 e^{-\frac{x^2 + z^2}{4\sigma_0^2}}$ and where $|\pm_{A}\rangle$ ($|\pm_{B}\rangle$) are the eigenvectors
of the spin operators $\widehat{s}_{z_A}$ ($\widehat{s}_{z_B}$) in
the z-direction pertaining to particule A (B): $\widehat{s}_{z_A}
|\pm_{A}\rangle= \pm (\frac{\hbar}{2})|\pm_{A}\rangle$
($\widehat{s}_{z_B} |\pm_{B}\rangle= \pm
(\frac{\hbar}{2})|\pm_{B}\rangle$). We treat classically
dependence with y: speed $- v_y$ for A and $ v_y$ for B.

The wave function $\Psi(\textbf{r}_A, \textbf{r}_B, t)$ of the two
identical particles A and B, electrically neutral and with
magnetic moments $\mu_0$, subject to magnetic fields
$\textbf{B}^{\mathcal{A}}$ and $\textbf{B}^{\mathcal{B}}$, admits
in the basis $|\pm_{A}\rangle$ and $|\pm_{B}\rangle$ 4 components
$\Psi^{a,b}(\textbf{r}_A, \textbf{r}_B, t)$ and verifies the
two-body Pauli equation~\cite{Holland_1993} p. 417:
\begin{eqnarray}
    i\hbar \frac{\partial \Psi^{a,b}}{\partial t}
    =\left(-\frac{\hbar^2}{2 m}\Delta_A -\frac{\hbar^2}{2 m}\Delta_B\right)\Psi^{a,b}
     +\mu B^{\mathcal{A}}_j (\sigma_j)_{c}^{a}\Psi^{c,b}
     +\mu B^{\mathcal{B}}_j (\sigma_j)_{d}^{b}\Psi^{a,d}\label{eq:7Paulideuxcorps1}
\end{eqnarray}
with the initial conditions:
\begin{equation}\label{eq:7Paulideuxcorps2}
\Psi^{a,b}(\textbf{r}_A, \textbf{r}_B,
0)=\Psi_0^{a,b}(\textbf{r}_A, \textbf{r}_B)
\end{equation}
where the $\sigma_j$ are the Pauli matrixes and where the
$\Psi_0^{a,b}(\textbf{r}_A, \textbf{r}_B)$ correspond to the
singlet state (\ref{eq:7psi-1}).

We take as numerical values those of the Stern-Gerlach experiment
with silver atoms~\cite{CohenTannoudji_1977,Gondran_2005b}. For a
silver atom, one has $m = 1,8\times 10^{-25}$ kg, $v_y = 500$\ m/s
, $\sigma_0$=10$^{-4}$m. For the electromagnetic field
$\textbf{B}$, $B_{x}=B'_{0} x$; $B_{y}=0$ and $B_{z}=B_{0} -B'_{0}
z$ with $B_{0}=5$ Tesla, $B'_{0}=\left| \frac{\partial B}{\partial
z}\right| =- \left| \frac{\partial B}{\partial x}\right|= 10^3$
Tesla/m over a length $\Delta l=1~cm$. The screen that intercepts
atoms is at a distance $D=20~cm$ (time $t_1=\frac{D}{v_y}= 4\times
10^{-4}$s) from the exit of the magnetic field.

One of the difficulties of the interpretation of the EPR-B
experiment is the existence of two simultaneous measurements. By
doing these measurements one after the other, the interpretation
of the experiment will be facilitated. That is the purpose of the
two-step version of the experiment EPR-B studied below.

\subsection{First step: Measurement of A spin and position~of~B}

In the first step we make, on a couple of particles A and B in a
singlet state, a Stern and Gerlach "measurement" for atom A, and
for atom B a mere impact measurement on a screen.

It is the experiment first proposed in 1987 by Dewdney, Holland
and Kyprianidis~\cite{Dewdney_1987b}.

Consider that at time $t_0$ the particle A arrives at the entrance
of electromagnet $\mathcal{A}$. $\triangle t$ is the crossing
duration of electromagnet $\mathcal{A}$ and $t$ is the time after
the $\mathcal{A}$ exit. The wave function can be calculated, from
the wave function (\ref{eq:7psi-1}), term to term in basis
[$|\pm_{A}\rangle,|\pm_{B}\rangle$]. After this exit of the
magnetic field $\mathcal{A}$, at time $t_0+ \triangle t + t$, the
wave function (\ref{eq:7psi-1}) becomes~\cite{Gondran_2005b}:
\begin{eqnarray}
\Psi(\textbf{r}_A, \textbf{r}_B, t_0 + \triangle t+ t )&=& \frac{1}
{\sqrt{2}}
f(\textbf{r}_B)\label{eq:7psiexperience1}\\
&&\times
\left( f^{+}(\textbf{r}_A,t) |+_{A}\rangle | -_{B}\rangle -
f^{-}(\textbf{r}_A,t) |-_{A}\rangle |
+_{B}\rangle\right)\nonumber
\end{eqnarray}
with
\begin{equation}\label{eq:7fonction}
f^{\pm}(\textbf{r},t)\simeq f(x, z \mp z_\triangle \mp ut)
e^{i(\frac{\pm muz}{\hbar}+ \varphi^\pm (t))}
\end{equation}
and
\begin{eqnarray}
    \Delta t= \frac{\Delta l}{v_y}=&&2 \times 10^{-5}s,~~~~z_{\Delta}=\frac{\mu_0 B'_{0}(\Delta
    t)^{2}}{2 m}=10^{-5}m,\nonumber\\
    &&u =\frac{\mu_0 B'_{0}(\Delta t)}{m}=1 m/s.\label{eq:7zdeltavitesse}
\end{eqnarray}
The atomic density $\rho(z_A, z_B,t_0 + \Delta t + t)$ is found by
integrating $\Psi^{*}(\textbf{r}_A,\textbf{r}_B, t_0 + \triangle t + t
)\Psi(\textbf{r}_A, \textbf{r}_B, t_0 + \triangle t+ t)$ on
$x_A$ and $x_B$:
\begin{eqnarray}
    \rho(z_A, z_B,t_0 + \Delta t+ t) &=& \left((2\pi\sigma_0^2)^{-\frac{1}{2}}
                  e^{-\frac{(z_B)^2}{2\sigma_0^2}}\right)
                  \label{eq:7densitéaprèschampmagnétiqueAB}\\
     &&\times\left((2\pi\sigma_0^2)^{-\frac{1}{2}}
                  \frac{1}{2}\left(e^{-\frac{(z_A-z_{\Delta}- ut)^2}{2\sigma_0^2}}+
                  e^{-\frac{(z_A+z_{\Delta}+
                  ut)^2}{2\sigma_0^2}}\right)\right).\nonumber
\end{eqnarray}

We deduce that the beam of particles A is divided into two, while
the B beam of particle stays one. This result can easily be tested
experimentally.

Moreover, we note that the space quantization of particle A is
identical to that of an untangled particle in a Stern and Gerlach
apparatus: the distance $\delta z= 2(z_{\Delta}+ ut)$ between the
two spots $N^+$ (spin +) and $N^-$ (spin $-$) of a family of
particle A is the same as the distance between the two spots $N^+$
and $N^-$ of a particle in a classic Stern and Gerlach
experiment~\cite{Gondran_2005b}. This result can easily be tested
experimentally.

We finally deduce from (\ref{eq:7densitéaprèschampmagnétiqueAB})
that:
\begin{itemize}
	\item the density of A is the same, whether particle A is entangled
with B or not,
	\item the density of B is not affected by the "measurement" of A.
\end{itemize}
These two predictions of quantum mechanics can be tested. Only
spins are involved. We conclude from (\ref{eq:7psiexperience1})
that the spins of A and B remain opposite throughout the
experiment.

\subsection{Second step: "Measurement" of A spin, then of B spin.}

The second step is a continuation of the first and results in
realizing the EPR-B experiment in two steps.

On a couple of particles A and B in a singlet state, first we made
a Stern and Gerlach "measurement" on the A atom between $t_0$ and
$t_0+ \triangle t+ t_1 $, then a  Stern and Gerlach "measurement"
on the B atom with an electromagnet $\mathcal{B}$ forming an angle
$\delta$ with $\mathcal{A}$ between $t_0 + \triangle t+ t_1$ and
$t_0+ 2( \triangle t + t_1)$.

Beyond the exit of magnetic field $\mathcal{A}$, at time $t_0 +
\triangle t +t_1$, the wave function is given by
(\ref{eq:7psiexperience1}). Immediately after the "measurement" of
A, still at time $t_0+ \triangle t+ t_1 $, if the A measurement is
$\pm$, the conditionnal wave functions of B are:
\begin{equation}\label{eq:7psiexperience1BcondmesA}
\Psi_{B /\pm A}(\textbf{r}_B, t_0 + \triangle t+ t_1 )=
f(\textbf{r}_B) |\mp_{B}\rangle.
\end{equation}
To measure B, we refer to the basis $|\pm'_{B}\rangle$ where
$|\pm'_{B}\rangle$ are the eigenvectors of the spin operators
$\widehat{s}_{z'_B}$ in the z'-direction pertaining to particule
B. We note $\textbf{r}'= (x',z')$. So, after the measurement of B,
at time $t_0+ 2( \triangle t +t_1)$ the conditional wave functions
of B are:
\begin{eqnarray}
\Psi_{B /+ A}(\textbf{r}'_B, t_0+ 2( \triangle t +t_1))=
\cos\frac{\delta}{2}f^{+}(\textbf{r}'_B,t_1)|+'_{B}\rangle
+\sin\frac{\delta}{2}f^{-}(\textbf{r}'_B,t_1)|-'_{B}\rangle,\label{eq:7psiexperience1B+A}
\\
\Psi_{B /- A}(\textbf{r}'_B, t_0+ 2(\triangle t +t_1))=
-\sin\frac{\delta}{2} f^{+}(\textbf{r}'_B,t_1)|+'_{B}\rangle +
\cos\frac{\delta}{2}
f^{-}(\textbf{r}'_B,t_1)|-'_{B}\rangle.\label{eq:7psiexperience1B-A}
\end{eqnarray}

We therefore obtain, in this two steps version of the EPR-B
experiment, the same results for spatial quantization and
correlations of spins as in the EPR-B experiment.

\section{Causal interpretation of the EPR-B experiment}

We assume, at moment of the creation of the two entangled
particles A and B, that each of the two particles A and B has an
initial wave function $\Psi_0^A(\textbf{r}_A, \theta^A_0,
\varphi^A_0)$ and $\Psi_0^B(\textbf{r}_B, \theta^B_0,
\varphi^B_0)$ with spinors which are opposite spins; for example
\newline
$\Psi_0^A(\textbf{r}_A, \theta^A_0, \varphi^A_0)= f(\textbf{r}_A)
\left(\cos\frac{\theta^A_0}{2}|+_{A}\rangle +
\sin\frac{\theta^A_0}{2}e^{i \varphi^A_0}|-_{A}\rangle\right)$ and
\newline
$\Psi_0^B(\textbf{r}_B , \theta^B_0, \varphi^B_0)= f(\textbf{r}_B)
\left(\cos\frac{\theta^B_0}{2}|+_{B}\rangle +
\sin\frac{\theta^B_0}{2}e^{i \varphi^B_0}|-_{B}\rangle\right)$ with
$\theta_0^B= \pi-\theta_0^A$ and $\varphi_0^B= \varphi_0^A -\pi$.

Then the Pauli principle tells us that the two-body wave function
must be antisymmetric; after calculation we find:
\begin{eqnarray*}
 \Psi_0(\textbf{r}_A,\theta^A, \varphi^A,\textbf{r}_B,\theta^B, \varphi^B)= - e^{i \varphi^A} f(\textbf{r}_A) f(\textbf{r}_B)
 \times
 \left(|+_{A}\rangle
|-_{B}\rangle - |-_{A}\rangle|+_{B}\rangle\right)
\end{eqnarray*}
which is the same as the singlet state, factor wise
(\ref{eq:7psi-1}).

Thus, we can consider that the singlet wave function is the wave
function of a family of two fermions A and B with opposite spins:
direction of initial spin A and B exist, but is not
\textit{known}. It is a local hidden variable which is therefore
necessary to add in the initial conditions of the model.

This is not the interpretation followed by the school of Bohm
~\cite{Dewdney_1987b,Holland_1988b,Holland_1993,Bohm_1993} in the
interpretation of the singlet wave function; they suppose, for
example, a zero spin for each of particles A and B at the initial
time.

It remains to determine the wave function and the trajectories of
particles A and B: from the entangled wave function, initial spins
and initial positions of each particle.

We assume therefore that the intial position of the particle A is
known ($x_0^A,y_0^A=0,z_0^A)$ as well as the particle B ($x_0^B =
x_0^A$,$y_0^B=y_0^A=0$,$z_0^B = z_0^A$).

\subsection{Step 1: Measurement of A spin and position of~B}

Equation (\ref{eq:7psiexperience1}) shows that the spins of A and
B remain opposite throughout step 1. Equation
(\ref{eq:7densitéaprèschampmagnétiqueAB}) shows that the densities
of A and B are independent; for A equal to the density of a family
of free particles in a classical Stern Gerlach apparatus, whose
initial spin orientation has been randomly chosen; for B equal to
the density of a family of free particles.

The spin of a particle A is orientated gradually following the
position of the particle in its wave into a spin $+$ or $-$. The
spin of particle B follows that of A, while remaining opposite.

In the equation (\ref{eq:7psiexperience1}) particle A can be
considerd independent of B. We can therefore give it the wave
function
\begin{eqnarray}
\Psi^A(\textbf{r}_A, t_0+ \triangle t+ t )=
\cos\frac{\theta_0^A}{2} f^{+}(\textbf{r}_A,t)|+_{A}\rangle
+ \sin\frac{\theta_0^A}{2}e^{i
\varphi_0^A}f^{-}(\textbf{r}_A,t)|-_{A}\rangle\label{eq:fonctiondondeA}
\end{eqnarray}
which is that of a free particle in a Stern Gerlach apparatus and
whose initial spin is given by ($\theta_0^A,\varphi_0^A$).

In de Broglie interpretation~\cite{Bohm_1993}, particle velocity is proportional to the gradient of the wave function phase.
See compute exemples for Young experiment~\cite{Gondran_2005a} and Stern-Gerlach experiment~\cite{Gondran_2005b}.
So, the equation of its trajectory is given by the following
differential equations: in the interval $[t_0,t_0 + \Delta t]$:
\begin{eqnarray}
 &&\frac{d z_A}{d t}=\frac{\mu_0 B'_{0} t}{ m} cos\theta(z_A,t)
 \nonumber\\
 &\text{with }&
 \tan \frac{\theta(z_A,t)}{2}=\tan\frac{\theta_0}{2} e^{- \frac{\mu_0 B'_{0} t^{2}z_A}
{2 m \sigma_0^{2}}}\label{eq:7trajectoiredanschampSetGA}
\end{eqnarray}
with the initial condition $z_A(t_0)=z^A_0$; and in the interval
$t_0 + \Delta t +t$ ($t\geq0$):
\begin{eqnarray}
 &&\frac{d z_A}{d t}=u \frac{\tanh(\frac{(z_\Delta + ut) z_A}{\sigma_{0}^{2}})+\cos \theta_0}{1+\tanh(\frac{(z_\Delta + ut) z_A}{\sigma_{0}^{2}})\cos
 \theta_0}
 \nonumber\\
 &\text{et }&\tan \frac{\theta(z_A(t),t)}{2}=\tan\frac{\theta_0}{2} e^{- \frac{(z_\Delta + ut)z_A}{\sigma_0^{2}}}.\label{eq:7trajectoireapreschampSetGA2}
\end{eqnarray}

$\theta(z_A(t),t)$ describes the evolution of the orientation of
spin A.

The case of particle B is different. B follows a rectilinear
trajectory with $y_B(t)= v_yt$, $z_B(t)=z_0^B$ and $x_B(t)=x_0^B$.
By contrast, the orientation of its spin moves and it was
$\theta^B(t)= \pi - \theta(z_A(t),t)$ and $\varphi^B(t)=
\varphi(z_A(t),t)- \pi$.

We can then associate the wave function:

\begin{eqnarray}
\Psi^B(\textbf{r}_B, t_0+ \triangle t+ t )=f(\textbf{r}_B)
\left(
\cos\frac{\theta^B(t)}{2} |+_{B}\rangle +
\sin\frac{\theta^B(t)}{2}e^{i \varphi^B(t)}|-_{B}\rangle\right).\label{eq:fonctiondondeB}
\end{eqnarray}

This wave funtion is specific, because it depends upon initial
conditions of A (positions and spins). The orientation of spin of
the particle B is driven by the particle A \textit{through the
singlet wave function}. Thus, the singlet wave function is the
actual non-local hidden variable.

\begin{figure*}
\begin{center}
\includegraphics[width=0.65\linewidth]{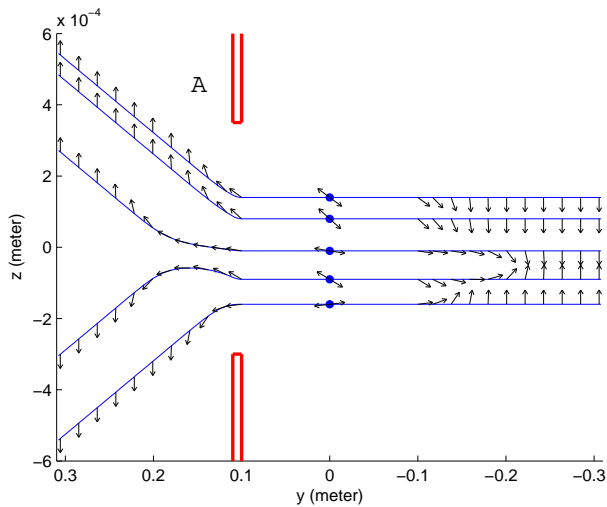}
\caption{\label{fig:EPR-5trajectoires}Five pairs of trajectories
of entangled particles. Arrows represent the spin orientation ($\theta$).}
\end{center}
\end{figure*}

Figure~\ref{fig:EPR-5trajectoires} presents a plot in the $(z,y)$
plane the trajectories of a set of 5 pairs of entangled atoms
whose initial characteristics $(\theta^A_0= \pi-
\theta^B_0,z_0^A=z_0^B)$ have been randomly chosen. The
trajectories will therefore depend on both the initial position
$z_0$ and the initial spin orientation $\theta_0$. Since the spin
initial orientation are different, trajectories of the A particles
may intersect.

\subsection{Step 2: "Measurement" of A spin, and then B spin}

Until time $t_0+ \triangle t+ t_1 $, we are in the case of step 1.
Immediately after the "measurement" of A at the time $t_0 + \Delta
t + t_1$, if the A measurement is $\pm$, the conditional wave
function of B is given by (\ref{eq:7psiexperience1BcondmesA}).

Then particle B is in position $(x_0^B, z_0^B)$.

We are exactly in the case of a particle in a Stern and Gerlach
magnet $\mathcal{B}$ which is an angle $\delta$ with
$\mathcal{A}$.

To measure the spin of B, we refer to the basis $|\pm'_{B}\rangle$. 
So, after the measurement of B, at time $t_0+ 2(
\triangle t +t_1)$, the conditional wave functions of B are given
by (\ref{eq:7psiexperience1B+A}) and
(\ref{eq:7psiexperience1B-A}), and we find again the quantum
correlations.

\section{Conclusion}

From the wave function of two entangled particles, we have
determined spins, trajectories and also a wave function for each
of the two particles.

In this interpretation, the quantum particle has a local position
like a classical particle, but it has also a non local behaviour
through the wave function. Indeed the wave function is not
separable and non-local. Because in the Broglie-Bohm
interpretation the wave function pilots the particle, it also
creates the non separability of two entangled particles.

As we saw in step 1, \textbf{the non-local influence in the EPR-B
experiment only concerns the spin orientation, and not the motion
of the particles themselves.} This is a key point in the search of
a physical explanation of non-local influence.

The simplest explanation (Ockham's razor) of this nonlocal
influence is to reintroduce the existence of a space having
certain properties related to the action at a distance, that is a
kind of ether, but a new form of ether given by Lorentz-Poincaré
and then by Einstein in 1920. Einstein said~\cite{Einstein_1920}:

"\textit{But on the other hand there is a weighty argument to be
adduced in favour of the ether hypothesis. To deny the ether is
ultimately to assume that empty space has no physical qualities
whatever. The fundamental facts of mechanics do not harmonize with
this view. For the mechanical behaviour of a corporeal system
hovering freely in empty space depends not only on relative
positions (distances) and relative velocities, but also on
\textbf{its state of rotation}, which physically may be taken as a
characteristic not appertaining to the system in itself. In order
to be able to look upon the rotation of the system, at least
formally, as something real, Newton objectivises space. Since he
classes his absolute space together with real things, for him
rotation relative to an absolute space is also something real.
Newton might no less well have called his absolute space "Ether";
\textbf{what is essential is merely that besides observable
objects, another thing, which is not perceptible, inust be looked
upon as real, to enable acceleration or rotation to be looked upon
as something real}.}[...]

\textit{Recapitulating, we may say that \textbf{according to the
general theory of relativity space is endowed with physical
qualities; in this sense, therefore, there exists an ether}.
According to the general theory of relativity \textbf{space
without ether is unthinkable}; for in such space there not only
would be no propagation of light, but also no possibility of
existence for standards of space and time (measuring-rods and
clocks), nor therefore any space-time intervals in the physical
sense. But this ether may not be thought of as endowed with the
quality characteristic of ponderable inedia, as consisting of
parts which may be tracked through time. \textbf{The idea of
motion may not be applied to it}.}"

Taking into account the new experiments, especially Aspect's
experiments, Popper~\cite{Popper_1982} (p. XVIII) defends a
similar view in 1982 :

"\textit{I feel not quite convinced that the experiments are
correctly interpreted; but if they are, we just have to accept
action at a distance. I think (with J.P. Vigier) that this would
of course be very important, but I do not for a moment think that
it would shake, or even touch, realism. Newton and Lorentz were
realists and accepted action at a distance; and Aspect's
experiments would be the first crucial experiment between
Lorentz's and Einstein's interpretation of the Lorentz
transformations.}"

Lastly, let us notice the great difference between EPR and EPR-B
experiments. The spin connected to the rotation of space-time
seems to be the cause of the instantaneous action at a distance in
experiment EPR-B. It is thus possible that there is not
instantaneous action at a distance in original experience EPR. And
in this case, Einstein was right. It is the proposal of
Popper~\cite{Popper_1982} p.25: " \textit{I mays perhaps mention
here some of the differences between the original EPR argument and
Bohm'version of it. These differences relate to the distinction of
two kinds of quantum mechanical state preparations.}" [...]
"\textit{Indeed, it is possible that the Bohm-Bell experiment
decides for action at a distance , and therefore against special
relativity theory, whereas the original EPR arguments does not}."

\textbf{The new experiments of non-locality have therefore a great
importance}, not to eliminate realism and determinism, but as
Popper said, \textbf{to rehabilitate the existence of a certain
type of ether}, like Lorentz's ether and like Einstein's ether in
1920.


\begin{thebibliography}{99}


\bibitem{EPR}
Einstein, A., Podolsky, B., Rosen,N.: Can quantum mechanical
description of reality be considered complete?. Phys. Rev.
47,777-780 (1935).


\bibitem{Bell64}
Bell,J. S.: On the Einstein Podolsky Rosen Paradox. Physics 1, 195
(1964).


\bibitem{CHSH}
Clauser,J.F., Horne,M.A., Shimony, A., Holt,R. A.:Proposed
experiments to test local hidden-variable theories. Phys. Rev.
Lett. 23, 880 (1969).


\bibitem{Bell_1987}
Bell,J. S.: Speakable and Unspeakable in Quantum Mechanics.
Cambridge University Press (1987).


\bibitem{Bohm_1951}
Bohm,D.:Quantum Theory. New York, Prentice-Hall (1951).


\bibitem{Bohm_1957}
Bohm,D., Aharonov,Y.: Discussion of experimental proofs for the
paradox of Einstein, Rosen and Podolsky. Phys. Rev.108, 1070
(1957).


\bibitem{Clauser_1972}
Freedman, S.J., Clauser,J.F.: Experimental test of local
hidden-variable theories. Phys. Rev. Lett. 28, 938 (1972).


\bibitem{Fry_1976}
Fry,E. S., Thompson,R.C.: Experimental Test of Local
Hidden-Variable Theories. Phys. Rev. Lett. 37, 465 (1976).


\bibitem{Lamehi_1976}
Lamehi-Rachti,M.,Mittig,W.: Phys. Rev. D 14, 2543(1976).


\bibitem{Aspect_1982a}
Aspect,A., Grangier,P., Roger,G.: Experimental realization of
Einstein-Pdolsky-Rosen-Bohm GedankenExperiment: a new violation of
Bell'inequalities. Phys. Rev. Lett. 49, 91 (1982).


\bibitem{Aspect_1982b}
Aspect,A., Dalibard, J., Roger,G.: Experimental tests of
Bell'inequalities using variable analysers. Phys. Rev. Lett. 49,
1804 (1982).


\bibitem{Tittel_1998}
Tittel,W., Brendel, J., Zbinden, H., Gisin,N.: Violation of Bell
inequalities by photons more than 10 km apart. Phys. Rev. Lett.
81, 3563 (1998).


\bibitem{Zeilinger_1998}
Weihs,G., Jennewein,T., Simon,C., Weinfurter,H., Zeilinger,A.:
Violation of Bell'inequalities under strict Einstein locality
condition. Phys. Rev. Lett. 81, 5039 (1998).


\bibitem{Beige}
Beige,A., Munro,W.J., Knight,P.L.: A Bell's inequality test with
entangled atoms. Phys. Rev. A 62, 052102-1-052102-9 (2000).


\bibitem{Rowe}
Rowe,M.A., Kielpinski,D., Meyer, V., Sackett,C.A., Itano,W.M.,
Monroe,C., Wineland,D.J.: Experimental violation of a Bell's
inequality with efficient detection. Nature 409, 791-794 (2001).


\bibitem{Zeilinger_2002}
Bertlmann,R.A., Zeilinger,A. (eds.): Quantum [un]speakables, from
Bell to Quantum information, Springer (2002).


\bibitem{Genovese_2005}
Genovese,M.: Research on hidden variables theories: a review of
recent progress. Phys. Repts. 413, 319 (2005).


\bibitem{CohenTannoudji_1977}
Cohen-Tannoudji,C., Diu,B., Laloë,F.: Quantum Mechanics, Wiley,
New York (1977).


\bibitem{Gondran_2005b}
Gondran,M., Gondran,A.: A complete analysis of the Stern-Gerlach
experiment using Pauli spinors. quant-ph/05 1276 (2005).


\bibitem{Gondran_2005a}
Gondran,M., Gondran,A.: Numerical simulation of the double-slit
interference with ultracold atoms. Am. J. Phys. 73, 6 (2005).


\bibitem{Dewdney_1987b}
Dewdney,C., Holland,P.R., Kyprianidis,A.: A causal account of
non-local Einstein-Podolsky-Rosen spin correlations. J. Phys. A:
Math. Gen. 20, 4717-32 (1987).


\bibitem{Holland_1988b}
Dewdney,C., Holland,P.R., Kyprianidis,A., Vigier,J.P.: Nature,
336, 536-44 (1988).


\bibitem{Bohm_1993}
Bohm,D., Hiley,B.J.: The Undivided Universe. Routledge, London and
New York (1993).


\bibitem{Holland_1993}
Holland, P.R.: The quantum Theory of Motion, Cambridge University
Press (1993).


\bibitem{Leggett_2003}
Leggett,A.: Nonlocal hidden-variable theories and quantum
mechanics: An incompatibility theorem. Found. Phys. 33, 1469-1493
(2003).


\bibitem{Zeilinger_2007}
S. Gröblacher, T. Paterek, R. Kaltenbaek, C. Brukner, M. Zukowski,
M. Aspelmeyer and A. Zeilinger, "An experimental test of non-local
realism", Nature, \textbf{446}, 871-875 (2007).


\bibitem{Einstein_1920}
A. Einstein, "Ether and the Theory of Relativity",
Einstein address delivered on May 5th, 1920, in the University of Leyden (1920).


\bibitem{Popper_1982}
K. Popper, Quantum Theory and the Schism in Physics: From the Postscript to The Logic of Scientific Discovery,
W. Bartley, III, Hutchinson, Londres (1982).


\end{thebibliography}
\end{document}